\documentclass[aps,prl,twocolumn,showpacs,floatfix,twoside]{revtex4-1}
\usepackage{amssymb}
\usepackage{graphicx}
\usepackage{amsmath}
\usepackage{color}
\usepackage{xcolor}
\usepackage{float}
\usepackage{subcaption}
\usepackage[justification=centerlast,font={small}]{caption}
\usepackage{natbib}

\begin{document}

\title{Thermodynamic quantum time crystals.}
\author{Konstantin B. Efetov}
\affiliation{Ruhr University Bochum, Faculty of Physics and Astronomy, Bochum, 44780,
Germany }
\affiliation{National University of Science and Technology \textquotedblleft
MISiS\textquotedblright, Moscow, 119049, Russia}
\affiliation{International Institute of Physics, UFRN, 59078-400 Natal, Brazil}
\date{\today }

\begin{abstract}
Although quantum time crystals have been proposed initially as macroscopic
and thermodynamically stable states, results of subsequent study seemed to
indicate that they could be realized only in systems out of equilibrium.
Here, investigating a rather general microscopic model we show that, in
contrast to the general belief, thermodynamically stable macroscopic quantum
time crystals can exist. The order parameter of this new state of matter is
periodic in both real and imaginary time but its average over the phase of
the oscillations equals zero. At the same time, correlation functions of
physical quantities at different times oscillate periodically in the
difference of the times without any decay, and this behavior can in
principle be observed experimentally.
\end{abstract}

\maketitle

Many materials have stable crystalline structures that are periodic in space
but not in time. Are thermodynamic states with a periodic time dependence of
physical quantities forbidden by fundamental laws of nature?

This question was raised by Wilczek \cite{wilczek} who proposed a concept of
quantum time crystals using a model that possessed a state with a current
oscillating in time. The work has attracted a great attention but a more
careful consideration of the model \cite{bruno} has led to the conclusion
that this was not an equilibrium state. These publications were followed by
a hot discussion of the possibility of realization of a thermodynamically
stable quantum time crystal \cite%
{wilczek1,li,bruno1,bruno2,nozieres,wilczek2}. More general arguments
against thermodynamically stable macroscopic quantum time crystals have been
put forward later \cite{watanabe}. As a result, a consensus has been
achieved that thermodynamically macroscopic quantum time crystals could not
exist although slowly decaying oscillations in systems out of equilibrium
were not forbidden. Recent theoretical \cite%
{volovik,sacha,sondhi1,sondhi2,nayak,yao} and experimental \cite%
{autti,zhang,choi} works have shown that this research field is interesting,
and at present, the term `Quantum Time Crystal' is used for non-equilibrium
systems.

Here, considering a model of interacting fermions it is demonstrated, that
the system can undergo a phase transition into a state with an order
parameter oscillating in both imaginary $\tau $ and real $t$ time. The
period of the oscillations in the imaginary time $\tau $ equals $\left(
mT\right) ^{-1},$ $T$ is temperature and $m$ is an integer, as required by
boundary conditions for boson fields. The periodic real time oscillations
can be observed in scattering cross sections or other quantities containing
correlation functions of two or more order parameters but there are no
oscillations in single time quantities. Thermodynamic quantum time crystal
(TQTC) proposed here can exist in an arbitrarily large volume and is a novel
type of ordered states of matter.

Although being rather general, the model considered here has been introduced
previously in a slightly different form of a spin-fermion model with
overlapping hot spots (SFMOHS) for description of underdoped superconducting
cuprates \cite{volkov1,volkov2,volkov3}. In the language of SFMOHS, the new
TQTC state is characterized by a loop currents order parameter oscillating
both in space and time. The phase of the oscillations in time is arbitrary,
and the integration over the phase, necessary within the computational
scheme, gives zero. As a result, the time reversal symmetry is broken but no
magnetic moments appear. These features may correspond to the still
mysterious pseudogap state \cite{timusk,norman,hashimoto}.

The model used here can be obtained e.g. by simplifying the Hamiltonian of
SFMOHS. Details of derivation as well as calculations, additional results,
and extended discussions are presented in Supplemental Material (SM) \cite%
{SM} and Ref. \cite{efetovPRB}. Hamiltonian $\hat{H}$ is introduced as
\begin{eqnarray}
\hat{H} &=&\sum_{p}c_{p}^{+}\left( \varepsilon _{p}^{+}+\varepsilon
_{p}^{-}\Sigma _{3}\right) c_{p}  \label{k1} \\
&&+\frac{1}{4V}\Big[\tilde{U}_{\mathrm{0}}\Big(\sum_{p}c_{p}^{+}\Sigma
_{1}c_{p}\Big)^{2}-U_{\mathrm{0}}\Big(\sum_{p}c_{p}^{+}\Sigma _{2}c_{p}\Big)%
^{2}\Big].  \notag
\end{eqnarray}%
Equation (\ref{k1}) describes interacting fermions of two bands $1$ and $2$
, and $p=\left\{ \mathbf{p,}\alpha \right\} $ stands for the momentum $%
\mathbf{\ p}$ and spin $\alpha .$ The energies $\varepsilon _{a}\left(
\mathbf{p}\right) $ are two-dimensional spectra in the bands counted from
the chemical potential $\mu ,$ $\varepsilon _{p}^{\pm }=\frac{1}{2}\left(
\varepsilon _{1}\left( \mathbf{p}\right) \pm \varepsilon _{2}\left( \mathbf{p%
}\right) \right) ,$ the interaction constants $U_{\mathrm{0}}$ and $\tilde{U}%
_{\mathrm{0}}$ are positive, while $V$ is the volume of the system.
Two-component vectors $c_{p}=\left\{ c_{p}^{1},c_{p}^{2}\right\} $ contain
creation and annihilation operators $c_{p}^{1}$ and $c_{p}^{2}$ for the
fermions of the bands $1$ and $2$. Matrices $\Sigma _{1},\Sigma _{2},\Sigma
_{3}$ are Pauli matrices in the space of numbers $1$ and $2$.

Hamiltonian $\hat{H}$ resembles the Bardeen-Cooper-Schrieffer (BCS)
Hamiltonian for Cooper pairs \cite{bcs} but contains a long-range
interaction of electron-hole pairs instead of the interaction of
electron-electron ones. Usually, such a form of the interaction makes
BCS-like mean field theories exact. An unusual feature of the Hamiltonian $%
\hat{H}$ is that it contains both the inter-band attraction (term with
matrix $\Sigma _{2}$) and repulsion (term with $\Sigma _{1}$). Neglecting
the repulsion leads in the language of SFMOHS to emergence of static loop
currents oscillating in space with the double period of the lattice \cite%
{volkov3}. This corresponds to a hypothetical d-density wave (DDW) state
\cite{chakravarty}. It is this repulsion term that can eventually make the
thermodynamic states with time-dependent oscillating correlation functions
energetically more favorable than conventional ones.

Time-dependent non-equilibrium solutions for the order parameter $\Delta
\left( t\right) $ in superconductors are known in many situations and are of
interest to both theory \cite%
{vk,spivak,barankov,altshuler,altshuler1,dzero,galitski,moor} and experiment
\cite{matsunaga,matsunaga1}. Perturbations of the order parameter $\delta
\Delta \left( t\right) $ oscillate in time and form slowly decaying
amplitude (Higgs) modes.

Here, it is shown that, in addition to DDW state, there is a region of
parameters where the system can undergo a phase transition into a state with
an order parameter oscillating in both real $t$ and imaginary time $\tau .$
A rather general proof of the `no-go' theorem of Ref. \cite{watanabe} misses
the possible existence of a time-dependent order parameter, and therefore
the conclusions of Ref. \cite{watanabe} cannot be used here (see \cite%
{efetovPRB} for more details). The solution $B\left( t\right) $ of equations
for the order parameter is not unique, and $B\left( t-t_{0}\right) $ is also
a solution for any $t_{0}.$ So, one should take into account this degeneracy
integrating over $t_{0}.$ Single-time physical quantities do not depend on
time after integration over $t_{0}$, and one can speak of thermodynamics.

Calculation of the partition function
\begin{equation}
Z=\mathrm{Tr}\exp \left( -\hat{H}/T\right)  \label{k2}
\end{equation}%
can be performed using the imaginary-time formalism with $\tau $ in the
interval $\left( 0,1/T\right) $ \cite{agd}. In order to compute trace `%
\textrm{Tr' }over all states of $\hat{H}$, it is convenient to first
decouple the quartic interaction terms of $\hat{H}$, Eq. (\ref{k1}), using
Gaussian integration over fields $b\left( \tau \right) $ and $b_{1}\left(
\tau \right) $ (Hubbard-Stratonovich transformation) satisfying boson
boundary conditions
\begin{equation}
b\left( \tau \right) =b\left( \tau +1/T\right) ,\quad b_{1}\left( \tau
\right) =b_{1}\left( \tau +1/T\right) .  \label{k5a}
\end{equation}%
\textrm{\ }This rather standard trick allows one to calculate exactly trace
in Eq. (\ref{k2}) and reduce calculation of $Z$ to computation of a
functional integral over the fields $b\left( \tau \right) $ and $b_{1}\left(
\tau \right) .$ The free energy $F=-T\ln Z$ takes the form
\begin{equation}
F=-T\ln \left[ \int \exp \left[ -\frac{\mathcal{F}\left[ b,b_{1}\right] }{T}%
\right] DbDb_{1}\right] .  \label{k7}
\end{equation}%
Free energy functional $\mathcal{F}\left[ b,b_{1}\right] $ entering Eq. (\ref%
{k7}) equals

\begin{eqnarray}
\frac{\mathcal{F}\left[ b,b_{1}\right] }{T} &=&\int_{0}^{1/T}\Big[-2\sum_{%
\mathbf{p}}\mathrm{tr}\left[ \ln H\left( \tau ,\mathbf{p}\right) \right]
_{\tau ,\tau }  \notag \\
&&+V\left( \frac{b^{2}\left( \tau \right) }{U_{\mathrm{0}}}+\frac{%
b_{1}^{2}\left( \tau \right) }{\tilde{U}_{\mathrm{0}}}\right) \Big]d\tau ,
\label{k10}
\end{eqnarray}%
In Eq. (\ref{k10}), `$\mathrm{tr}$' means trace in space of the bands $1,2,$
and $H\left( \tau ,\mathbf{p}\right) =h-b_{1}\left( \tau \right) \Sigma
_{1}, $ where
\begin{equation}
h\left( \tau ,\mathbf{p}\right) =\partial _{\tau }+\varepsilon ^{+}\left(
\mathbf{p}\right) -\varepsilon ^{-}\left( \mathbf{p}\right) \Sigma
_{2}-b\left( \tau \right) \Sigma _{3}.  \label{k6}
\end{equation}%
The fields $b\left( \tau \right) $, $b_{1}\left( \tau \right) $ in Eqs. (\ref%
{k5a}-\ref{k6}) depend on the time $\tau $ only, which is a consequence of a
special form of the interaction in Eq. (\ref{k1}). The functional $\mathcal{F%
}\left[ b,b_{1}\right] $ is proportional to the volume $V,$ and in the limit
$V\rightarrow \infty $ the computation of the functional integral over $%
b\left( \tau \right) $ and $b_{1}\left( \tau \right) $ in Eq. (\ref{k7})
reduces to replacing $\mathcal{F}\left[ b,b_{1}\right] $ by its value at
minimum (saddle point approximation). This approximation is equivalent to
using the mean field theory that becomes exact in the limit $V\rightarrow
\infty .$

Minimizing $\mathcal{F}\left[ b,b_{1}\right] $ with respect to $b\left( \tau
\right) $, $b_{1}\left( \tau \right) $ we obtain the following equations
\begin{eqnarray}
b\left( \tau \right) &=&-U_{\mathrm{0}}\mathrm{tr}\int \Sigma _{3}\left[
H^{-1}\left( \tau ,\mathbf{p}\right) \right] _{\tau ,\tau }\frac{d\mathbf{p}%
}{\left( 2\pi \right) ^{2}},  \label{k3} \\
b_{1}\left( \tau \right) &=&-i\tilde{U}_{\mathrm{0}}\mathrm{tr}\int \Sigma
_{1}\left[ H^{-1}\left( \tau ,\mathbf{p}\right) \right] _{\tau ,\tau }\frac{d%
\mathbf{p}}{\left( 2\pi \right) ^{2}}.  \label{k3a}
\end{eqnarray}%
Although the minimum at
\begin{equation}
b\left( \tau \right) =\gamma ,\;b_{1}\left( \tau \right) =0  \label{k8}
\end{equation}%
obtained previously \cite{volkov3} is a minimum of $\mathcal{F}\left[ b,b_{1}%
\right] $ and Eq. (\ref{k8}) is a solution of Eqs. (\ref{k3}, \ref{k3a}), we
show that there is a region of parameters where the absolute minimum is
reached at $\tau $-dependent functions $b\left( \tau \right) $ and $%
b_{1}\left( \tau \right) .$

Now we sketch the main steps of the calculation of the free energy $F$ (see
also Refs. \cite{SM,efetovPRB}). The real-time behavior of correlation
functions at $T=0$ will be studied using a Wick rotation $\tau \rightarrow
it.$

We start with considering all possible extrema of $\mathcal{F}\left[ b(\tau
),0\right] $, Eq. (\ref{k10}). Varying the functional $\mathcal{F}\left[
b\left( \tau \right) ,0\right] $ one obtains
\begin{equation}
b\left( \tau \right) =-U_{0}\mathrm{tr}\int \Sigma _{3}\left[ h^{-1}\left(
\tau ,\mathbf{p}\right) \right] _{\tau ,\tau }\frac{d\mathbf{p}}{\left( 2\pi
\right) ^{2}}.  \label{k10b}
\end{equation}%
Actually, Eq. (\ref{k10b}) determines minima of $\mathcal{F}\left[ b,b_{1}%
\right] $, Eq. (\ref{k10}), exactly at $\tilde{U}_{\mathrm{0}}=0$. Although
Eq. (\ref{k10b}) is still quite non-trivial due to a possible dependence of $%
b\left( \tau \right) $ on $\tau $, its solutions $b_{0}\left( \tau \right) $
can be written exactly in terms of a Jacobi double-periodic elliptic
function $\mathrm{sn}\left( x|k\right) $,
\begin{equation}
b_{0}\left( \tau \right) =k\gamma \mathrm{sn}\left( \gamma \left( \tau -\tau
_{0}\right) |k\right) ,  \label{k12}
\end{equation}%
where parameter $k,$ $0<k<1,$ is the modulus, $\gamma $ is an energy, and $%
\tau _{0}$ is an arbitrary shift of the imaginary time in the interval $%
0<\tau _{0}<1/T$ . In the limit $k\rightarrow 1$, the function has an
asymptotic behavior $\mathrm{sn}\left( x|k\right) \rightarrow \pm \tanh x,$
while in the limit $k\ll 1$ one obtains $\mathrm{sn}\left( x|k\right)
\rightarrow \sin x.$

The period of the oscillations for an arbitrary $k$ equals $4K\left(
k\right) /\gamma ,$ where $K\left( k\right) $ is the elliptic integral of
the first kind, and therefore the condition
\begin{equation}
\gamma =4K\left( k\right) mT  \label{k12a}
\end{equation}%
with integer $m$, must be satisfied to fulfill Eqs. (\ref{k5a}). In the most
interesting limit of small $1-k,$ the period $4K\left( k\right) /\gamma $ of
$b_{0}\left( \tau \right) $ grows logarithmically as $\frac{1}{2}\ln $ $%
\left( \frac{8}{1-k}\right) ,$ and the solution $b_{0}\left( \tau \right) $
consists of $2m$ well separated alternating instantons and anti-instantons
with the shape $\pm \gamma \tanh \gamma \tau $. The integral over the period
of the oscillations in Eq. (\ref{k12}) as well as over the position $\tau
_{0}$ of the instanton equals zero.

Existence of the non-trivial local minima (\ref{k12}) of $\mathcal{F}\left[
b,0\right] $ at $b_{0}\left( \tau \right) $ has been established previously
by Mukhin \cite{mukhin,mukhin1} starting from a different model. Generally,
there can be many solutions corresponding to different minima of $\mathcal{F}%
\left[ b,0\right] $ depending on the number $m$ of instanton-antiinstanton
pairs (IAP). However, the lowest value of the functional $\mathcal{F}\left[
b,0\right] $ is reached at $m=0$ corresponding to the static order (see for
details Refs. \cite{efetovPRB,mukhin2}).

Presence of $b_{1}\left( \tau \right) $ in Eqs. (\ref{k10}, \ref{k3}, \ref%
{k3a}) can change the situation and this can be seen from the first order $%
\mathcal{F}_{1}\left[ b_{0},b_{1}\right] $ of expansion of the first term of
the free energy functional $\mathcal{F}\left[ b,b_{1}\right] $, Eq. (\ref%
{k10}),
\begin{equation}
\frac{\mathcal{F}_{1}\left[ b_{0},b_{1}\right] }{VT}=-2J\int_{0}^{1/T}\dot{b}%
_{0}\left( \tau \right) b_{1}\left( \tau \right) d\tau ,  \label{k11}
\end{equation}%
where $J$ is a constant (see \cite{SM}, \cite{efetovPRB}) depending on the
parameters of the Hamiltonian $\hat{H}$, Eq. (\ref{k1}). Equation (\ref{k11}%
) demonstrates that the field $b_{1}\left( \tau \right) $ linearly couples
to the time derivative of $b_{0}\left( \tau \right) ,$ Eq. (\ref{k12}). The
functional $\mathcal{F}_{1}\left[ b_{0},b_{1}\right] $ is real for positive $%
\tilde{U}_{\mathrm{0}}$, and this linear coupling can destabilize the static
minimum, Eq. (\ref{k8}). This is the key finding of the present work.
Replacing $\mathcal{F}\left[ b,b_{1}\right] $, Eq. (\ref{k10}), by
\begin{equation}
\mathcal{\tilde{F}}\left[ b,b_{1}\right] \approx \mathcal{F}\left[ b_{0},0%
\right] +\mathcal{F}_{1}\left[ b_{0},b_{1}\right] +VT\int_{0}^{1/T}\frac{%
b_{1}^{2}\left( \tau \right) }{\tilde{U}_{\mathrm{0}}},  \label{k13}
\end{equation}%
and minimizing $\mathcal{\tilde{F}}\left[ b,b_{1}\right] $ with respect to $%
b_{1}\left( \tau \right) $ one obtains an effective `instanton-instanton
attraction' described by the negative contribution $\mathcal{F}_{\mathrm{II}}%
\left[ b_{0},0\right] $,
\begin{equation}
b_{1}\left( \tau \right) =J\tilde{U}_{\mathrm{0}}\dot{b}_{0}\left( \tau
\right) ;\quad \mathcal{F}_{\mathrm{II}}\left[ b_{0},0\right] =-\tilde{U}_{%
\mathrm{0}}J^{2}\int_{0}^{1/T}\dot{b}_{0}^{2}\left( \tau \right) d\tau ,
\label{k13a}
\end{equation}%
The functional $\mathcal{F}_{\mathrm{II}}\left[ b_{0},0\right] $  should be
added to $\mathcal{F}\left[ b_{0},0\right] ,$ which gives the free energy $%
\mathcal{\tilde{F}}\left[ b_{0}\right] $ at the new minimum
\begin{equation}
\mathcal{\tilde{F}}\left[ b_{0}\right] =\mathcal{F}\left[ b_{0},0\right] +%
\mathcal{F}_{\mathrm{II}}\left[ b_{0},0\right] .  \label{k13b}
\end{equation}%
Presence of the negative term $\mathcal{F}_{\mathrm{II}}\left[ b_{0},0\right]
$ favors formation of $\tau $ -dependent structures.

We simplify our study by considering the limit of low temperatures $T$ when
one can expect a large number of IAP in the system and of small $1-k$
corresponding to a large period of the IAP lattice. In this limit, the
difference $\Delta F$ between the total free energy $F$ and the free energy $%
F_{\mathrm{st}}$ of the system with the static order parameter is
proportional to $2m.$ The case $\Delta F/\left( TV\right) >0$ corresponds to
the state with the static order, while in the region of parameters where $%
\Delta F/\left( TV\right) <0$ one can expect a chain of alternating
instantons and anti-instantons.

In the limit $k\rightarrow 1$, one can write for $\Delta F/V\left(
2mT\right) $ using Eqs. (\ref{k10}, \ref{k13}, \ref{k13a})
\begin{equation}
\Delta F=F_{\mathrm{inst}}+F_{\mathrm{II}}.  \label{k14}
\end{equation}%
In Eq. (\ref{k14}), $F_{\mathrm{inst}}$ is the energy of `non-interacting'
instantons originating from $\mathcal{F}\left[ b,0\right] $,
\begin{equation*}
\frac{F_{\mathrm{inst}}}{2mVT}=\int \left[ \ln \frac{E\left( \mathbf{p}%
\right) +\gamma }{E\left( \mathbf{p}\right) -\gamma }-\frac{2\gamma }{%
E\left( \mathbf{p}\right) }\right] \frac{d\mathbf{p}}{\left( 2\pi \right)
^{2}},
\end{equation*}%
while the contribution of the `instanton-instanton interaction' coming from $%
\mathcal{F}_{1}\left[ b_{0},b_{1}\right] $ equals%
\begin{equation*}
\frac{F_{\mathrm{II}}}{2mVT}=-\frac{\tilde{U}_{\mathrm{0}}}{4}\left[ \int
\frac{sgn\left( \varepsilon ^{-}\left( \mathbf{p}\right) \right) }{E\left(
\mathbf{p}\right) \sqrt{\left( \left( \varepsilon ^{-}\left( \mathbf{p}%
\right) \right) ^{2}+\frac{\gamma ^{2}\left( 1-k\right) ^{2}}{4}\right) }}%
\frac{d\mathbf{p}}{\left( 2\pi \right) ^{2}}\right] ^{2},
\end{equation*}%
where
\begin{equation*}
E\left( \mathbf{p}\right) =\sqrt{\left( \varepsilon ^{-}\left( \mathbf{p}%
\right) \right) ^{2}+\gamma ^{2}}.
\end{equation*}%
The energy $\gamma $ plays a role of the gap in the spectrum and can be
found from the static solution of Eq. (\ref{k10b}). This equation reduces in
the limit $k\rightarrow 1$ to the form
\begin{equation}
1=U_{0}\int_{0}^{1/T}\frac{1}{E\left( \mathbf{p}\right) }\frac{d\mathbf{p}}{%
\left( 2\pi \right) ^{2}},  \label{k21}
\end{equation}%
which is actually the equation for the order parameter of the DDW \cite%
{volkov3}.

\begin{figure}[tph]
\centering
\begin{subfigure}[b] {0.49\columnwidth}
  \includegraphics[width=\linewidth]{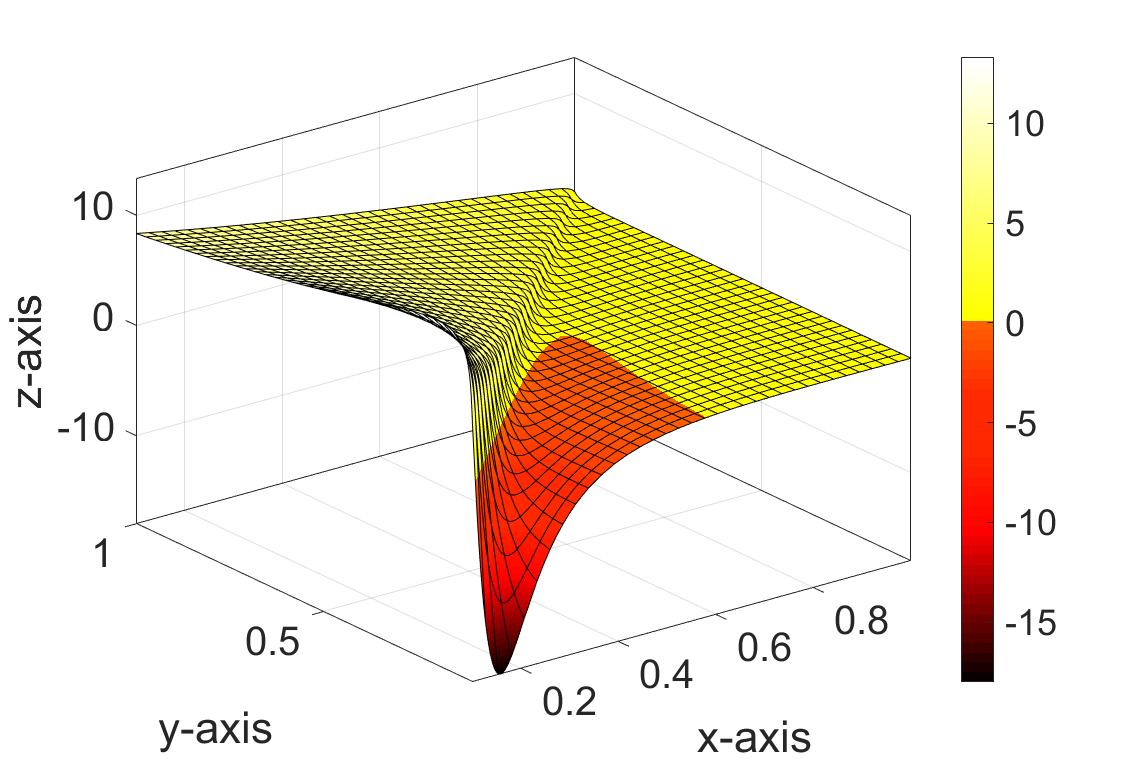}
 \caption{$k=0.99$, $a=0$}
 \end{subfigure}
\hfill
\begin{subfigure}[b]{0.49\columnwidth}
 \includegraphics[width=\linewidth]{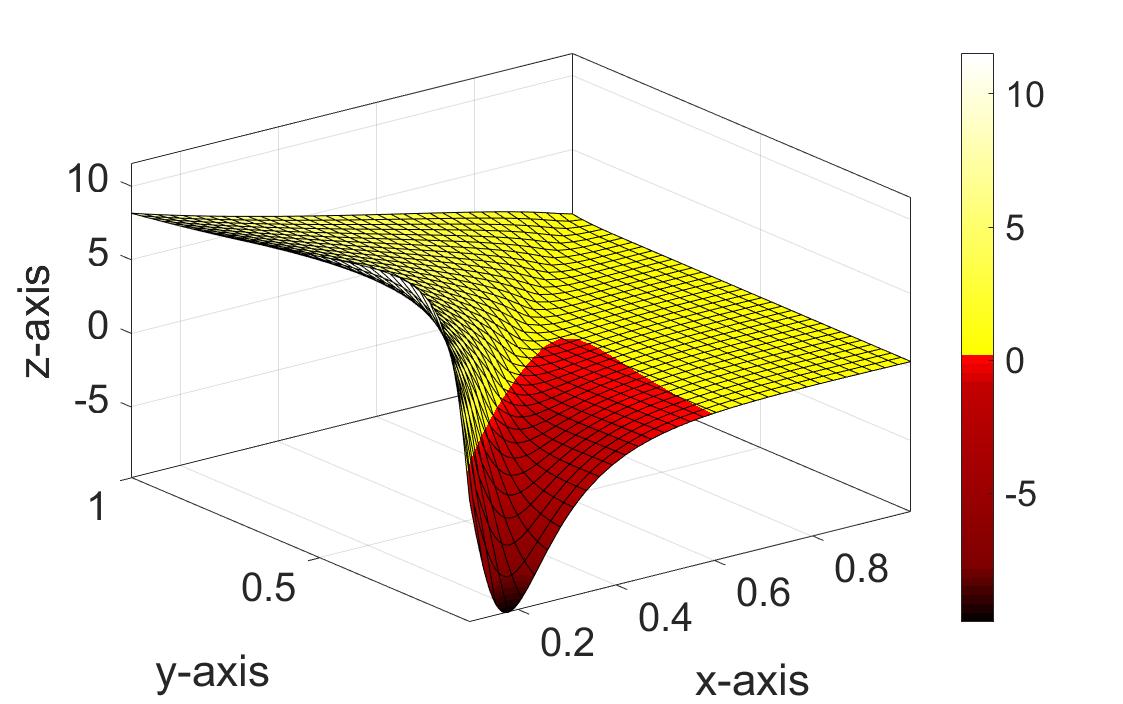}
 \caption{$k=0.90$, $a=0$}
 \end{subfigure}
\par
\begin{subfigure}[b]{0.49\columnwidth}
 \includegraphics[width=\linewidth]{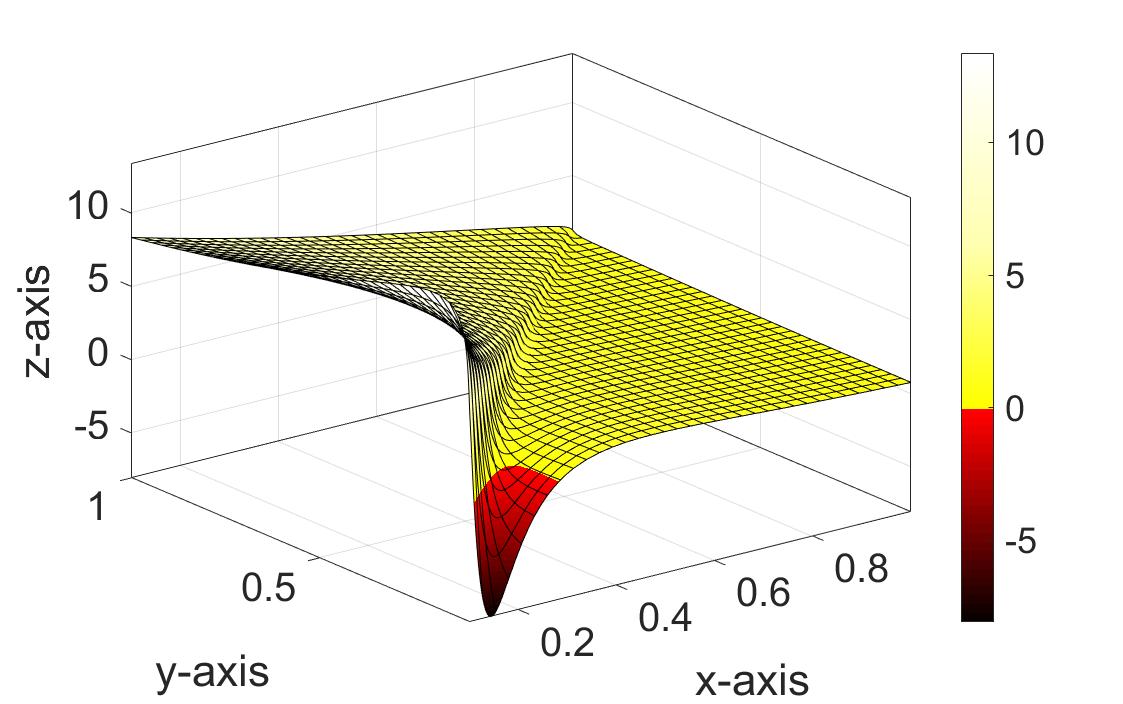}
 \caption{$k=0.99$, $a=1$}
 \end{subfigure}
\hfill
\begin{subfigure}[b]{0.49\columnwidth}
 \includegraphics[width=\linewidth]{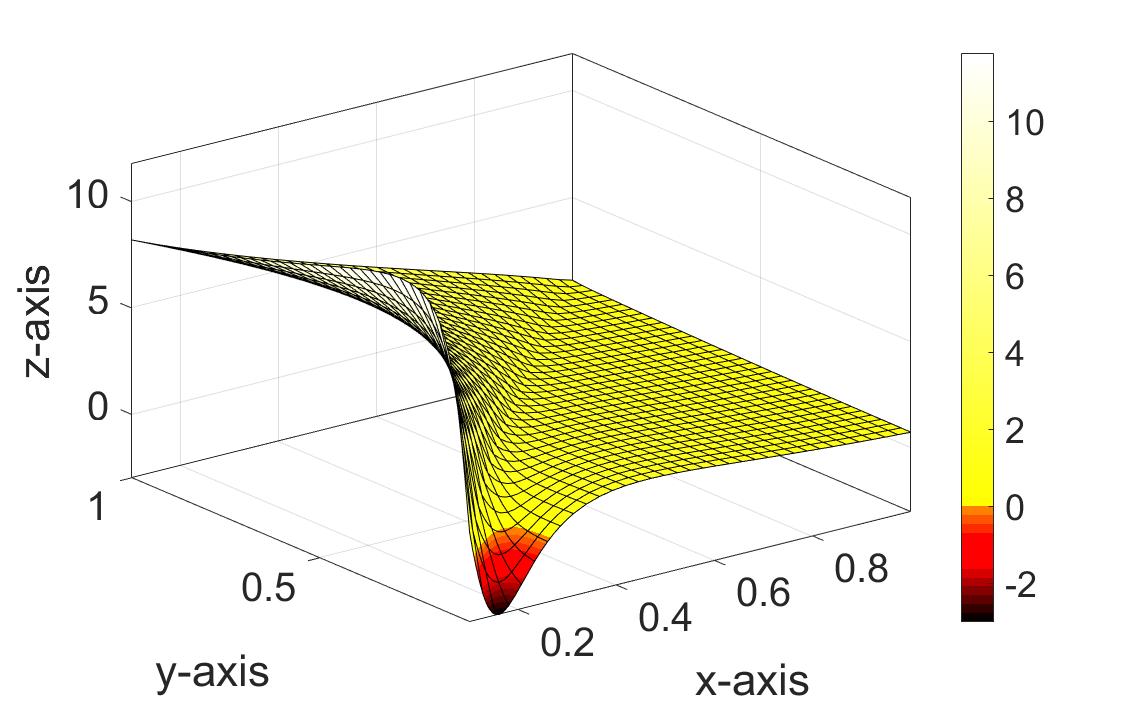}
 \caption{$k=0.90$, $a=1$}
 \end{subfigure}
\caption{(Color online.) Free energy of instanton-antiinstanton pairs.}
\label{fig:energy}
\end{figure}

The energy $F_{\mathrm{inst}}$ is always positive but $F_{\mathrm{II}}$ is
negative. It is important that in the limit $k\rightarrow 1$ the term $F_{%
\mathrm{II}}$ can be very large due to contributions coming from the region
of very small $\left\vert \varepsilon ^{-}\left( \mathbf{p}\right)
\right\vert $ thus making $\Delta F$ negative$.$ Moreover, the interaction $%
\tilde{U}_{\mathrm{0}}$ can considerably exceed $U_{\mathrm{0}}$ increasing
the negative contribution. Taking into account quadratic terms of the
expansion of $\mathcal{F}\left[ b,b_{1}\right] $ in $b_{1}$ leads to a
screening of the `inter-instanton interaction' and it is also taken into
account in Fig. \ref{fig:energy}(a-d).

The dependence of $S=\Delta F/\left( 2mTV\right) $ on parameters
characterizing the energy spectrum in SFMOHS is represented in Figs. \ref%
{fig:energy}(a-d). Computation of the free energy is performed choosing
\begin{equation*}
\varepsilon _{1}\left( \mathbf{p}\right) =\alpha p_{x}^{2}-\beta
p_{y}^{2}+P-\mu ,\quad \varepsilon _{2}\left( \mathbf{p}\right) =\alpha
p_{y}^{2}-\beta p_{x}^{2}-P-\mu
\end{equation*}%
corresponding to the spectrum of cuprates near the middle of the edges of
the Brillouin zone (momenta $\mathbf{p}$ are counted from the middle of the
edges), where $P$ is a Pomeranchuk order parameter obtained previously in
SFMOHS \cite{volkov1}, and $\mu $ is the chemical potential. We use a
parameter $a=U_{\mathrm{0}}/\tilde{U}_{\mathrm{0}},$ and an energy cutoff $%
\Lambda $ determining the boundary of the hot spots,
\begin{equation*}
a=U_{0}/\tilde{U}_{0},\quad \quad \left( \alpha +\beta \right) \left(
p_{x}^{2}+p_{y}^{2}\right) /2<\Lambda .
\end{equation*}%
Figs. \ref{fig:energy} show dependence of $z=2\pi ^{2}\left( \alpha +\beta
\right) S/\Lambda $ on $x=P/\gamma $ and $y=\Lambda /\gamma $,
and demonstrate existence of a region of parameters where the free energy $%
\Delta F$ is negative and the static state is unstable. The region of small $%
1-k$ and $a$ is most favorable for the formation of the lattice of IAP. As
we consider here structures periodic in space (oscillations with vector $%
\mathbf{Q}_{AF}$ connecting the bands $1$ and $2$), the periodic in $\tau $
order parameter $b\left( \tau \right) $ providing the minimum of the free
energy is at the same time the amplitude of the periodic oscillations in
space. The present consideration does not determine the number of the pairs $%
m$ as a function of temperature, and a more accurate study remains to be
performed in the future. Below we calculate physical quantities without
specifying the value of $m$.

The periodic structure described by the Jacobi elliptic function $%
b_{0}\left( \tau \right) $ (\ref{k12}) is actually double periodic in the
complex plane of $\tau $ and, hence, is periodic in real time $t.$
Remarkably, $b_{0}\left( it\right) $ still satisfies Eq. (\ref{k10b}) after
the rotation $\tau \rightarrow it.$ Generally, real-time correlation
functions can be calculated using a similar technique as previously. At $T=0$
one should replace in Eq. (\ref{k10}) $\tau \rightarrow it$ and integrate
over $t$ from $-\infty $ to $\infty .$ Repeating the steps made within the
imaginary-time representation one should integrate over functions $B\left(
t\right) ,$ $B_{1}\left( t\right) ,$ instead of $b\left( \tau \right) ,$ $%
b_{1}\left( \tau \right) .$ Proceeding in this way one obtains formulas
similar to Eqs. (\ref{k10}, \ref{k3}, \ref{k3a}) but written in real time $%
t. $ One can prove that the order parameters $B\left( t\right) $ and $%
B_{1}\left( t\right) $ are related to $b\left( \tau \right) $ and $%
b_{1}\left( \tau \right) $ as \cite{SM,efetovPRB}
\begin{equation}
iB\left( t\right) =b\left( it\right) ,\quad B_{1}\left( t\right)
=b_{1}\left( it\right) .  \label{k15}
\end{equation}%
If $B\left( t\right) $ and $B_{1}\left( t\right) $ provide the extremum of
the action, the same do $B\left( t-t_{0}\right) $ and $B_{1}\left(
t-t_{0}\right) $ for an arbitrary shift $t_{0}.$ Due to this degeneracy one
should average over $t_{0}$ at the end of calculations. Physically relevant
correlations of the loop currents in the model considered are described
exactly by $2$-times correlation function
\begin{equation}
N\left( t\right) =\frac{U_{0}^{2}}{V^{2}}\sum_{p,p^{\prime }}\left\langle
\left( c_{p}^{+}\left( t\right) \Sigma _{2}c_{p}\left( t\right) \right)
\left( c_{p^{\prime }}^{+}\left( 0\right) \Sigma _{2}c_{p^{\prime }}\left(
0\right) \right) \right\rangle .  \label{k17}
\end{equation}%
Using the saddle point equations (\ref{k3}, \ref{k3a}), replacing $\tau
\rightarrow it$, and using Eq. (\ref{k15}) we obtain in the limit $%
V\rightarrow \infty $
\begin{equation}
N\left( t\right) =\overline{B\left( t-t_{0}\right) B\left( -t_{0}\right) },
\label{k18}
\end{equation}%
where the bar stands for averaging over $t_{0}.$ One can generally expect a
periodic time-dependence of $N\left( t\right) $ using exact solutions for $%
B\left( t\right) .$ It is not easy to find them analytically but we show in
Sec. V of Ref. \cite{efetovPRB} that they are periodic in time, which
guarantees the periodicity of the function $N\left( t\right) .$ The averaged
order parameter vanishes
\begin{equation}
\overline{B\left( t-t_{0}\right) }=0.  \label{k16a}
\end{equation}

Now we approximate, as it was done previously, the function $B\left(
t\right) $ by a function $B_{0}\left( t\right) =-ib_{0}\left( it\right) $
and use Eq. (\ref{k12}). The Jacobi elliptic function $\mathrm{sn}\left(
iu,k\right) $ of an imaginary argument $iu$ is related to an elliptic
function $\mathrm{sc}\left( u|k\right) $ with the period $2K\left( k\right) $
\cite{as}
\begin{equation*}
\mathrm{sn}\left( iu|k\right) =i\mathrm{sc}\left( u|k^{\prime }\right)
,\;k^{2}+k^{\prime 2}=1,
\end{equation*}%
and we write $B\left( t\right) $ in Eq. (\ref{k18}) in the form
\begin{equation}
B\left( t\right) \approx B_{0}\left( t\right) =\gamma k\mathrm{sc}\left(
\gamma \left( t-t_{0}\right) |k^{\prime }\right) ,  \label{k16}
\end{equation}%
Function $N\left( t\right) $, Eq. (\ref{k18}), can be calculated using a
Fourier series for the function $\mathrm{sc}\left( u|k\right) $ \cite{as}.
Integration over $t_{0}$ gives in the limit $1-k\ll 1$ $1-k\ll 1$ \cite%
{SM,efetovPRB}
\begin{equation*}
N\left( t\right) \approx 2\gamma ^{2}\sum_{n=1}^{\infty }f_{n}^{2}\cos
\left( 2\gamma nt\right) ,\quad f_{n}=\left[ 1-\frac{1}{2}\left( \frac{1-k}{8%
}\right) ^{2n}\right] .
\end{equation*}

Function $N\left( t\right) $ shows an oscillating behavior with the
frequencies $2\gamma n$. The energy $2\gamma $ is the energy of the breaking
of electron-hole pairs, and one can interpret the form of $N\left( t\right) $
as oscillations between the static order and normal state. The oscillations
of $N\left( t_{1}-t_{2}\right) $ resemble those of the order parameter in
the non-equilibrium superconductors \cite%
{vk,spivak,barankov,altshuler,altshuler1,dzero,galitski,moor} but now the
function $N\left( t\right) $ does not decay in time. The contribution of
high harmonics $n$ does not decay with $n$ but apparently this is a
consequence of the approximation (\ref{k16}) for $B\left( t\right) $. At the
same time, one can generally expect a periodic time-dependence of $N\left(
t\right) $ using exact solutions for $B\left( t\right) .$ It is not easy to
find them analytically but one can show that they are periodic in time (Sec.
V of Ref. \cite{efetovPRB}), which guarantees the periodicity of the
function $N\left( t\right) .$

The order parameter $B\left( t-t_{0}\right) $ appears when calculating
fermionic quantum averages corresponding to the loop currents, and Eq. (\ref%
{k16a}) demonstrates that physical currents are equal to zero at any time $t$%
. Non-vanishing oscillations of two-times correlation function $N\left(
t\right) $ allow us to classify the physical state found here as
thermodynamic quantum time-space crystal.

The correlation function $N\left( t\right) $, Eq. (\ref{k18}), was
calculated integrating over the position $t_{0}.$ Remarkably, the same
results for correlation functions can be obtained considering a Hamiltonian $%
\hat{H}_{TC}$ \textrm{\ }of a harmonic oscillator
\begin{equation}
\hat{H}_{TC}=2\gamma \left( a^{+}a+1/2\right) ,  \label{k20}
\end{equation}%
where $a^{+}$ and $a$ are boson creation and annihilation operators (for
simplicity, we consider here the limit $1-k\ll 1$), and an `operator order
parameter' $A.$ Using the Hamiltonian $\hat{H}_{TC}$ one can write the
correlation function $N\left( t_{1}-t_{2}\right) $ in the form
\begin{equation*}
N\left( t\right) =\gamma ^{2}\left( \left\langle 0\left\vert A\left(
t\right) A^{+}\left( 0\right) \right\vert 0\right\rangle +\left\langle
0\left\vert A\left( 0\right) A^{+}\left( t\right) \right\vert 0\right\rangle
\right) ,
\end{equation*}%
where
\begin{equation*}
A^{+}\left( t\right) =e^{i\hat{H}_{TC}t}A^{+}e^{-i\hat{H}_{TC}t},\;A^{+}=%
\sum_{n=1}^{\infty }f_{n}\frac{\left( a^{+}\right) ^{n}}{\sqrt{n!}}
\end{equation*}%
and $\left\vert 0\right\rangle $ stands for the wave function of the ground
state of the Hamiltonian $\hat{H}_{TC}$ (\ref{k20}). At the same time,
quantum averages of the operators $A$ and $A^{+}$ vanish
\begin{equation*}
\left\langle 0\left\vert A\left( t\right) \right\vert 0\right\rangle
=\left\langle 0\left\vert A^{+}\left( t\right) \right\vert 0\right\rangle =0.
\end{equation*}%
The operator order parameters extends the variety of conventional order
parameters like scalars, vectors, matrices used in theoretical physics. The
non-decaying time oscillations is an important property for designing qubits.

Possibility of an experimental observation depends on systems described by
the Hamiltonian (\ref{k1}). For cuprates, inelastic polarized neutron
spectroscopy can be a proper tool for observations. Calculating the Fourier
transform $N\left( \omega \right) $ of the function $N\left( t\right) $ and
comparing it with the one for the hypothetical time-independent DDW result $%
2\pi \gamma ^{2}\delta \left( \omega \right) $, one can write at low
temperatures the ratio of the responses at $\left( \pi ,\pi \right) $ for
these two states as
\begin{equation}
\chi \left( \omega ,\mathbf{q}\right) =\chi _{0}\sum_{n=1}^{\infty
}f_{n}\delta \left( \omega -2n\gamma \right) \delta \left( \mathbf{q-Q}%
_{AF}\right) ,  \label{k23a}
\end{equation}%
where $\chi _{0}$ determines the response $\chi _{DDW}$ of the DDW state, $%
\chi _{DDW}\left( \omega \right) =\chi _{0}\delta \left( \omega \right) .$
According to Eq. (\ref{k23a}) elastic scattering cannot lead to any signal
expected for DDW. Actually, anisotropic magnetic $\left( \pi ,\pi \right) $
excitations have been observed \cite{hayden} in $YBa_{2}Cu_{3}O_{6.9}$ but
more detailed experiments are necessary to clarify their origin.

The main conclusion of the present study is that the quantum time-space
crystals may exist as a thermodynamically stable state in macroscopic
systems. The order parameter of TQTC is periodic in both real and imaginary
times but its average over the phase of the oscillations vanishes. The
non-decaying oscillations can be seen, e.g., in two- or more times
correlation functions. This leads to a natural generalization of the notion
of the space long-range order to the time-space one. Two-times correlation
functions determine cross-section in inelastic scattering experiments. The
frequency of the oscillations remains finite in the limit of infinite
volume, $V\rightarrow \infty $. One can expect various experimental
consequences and, in particular, one can suppose that the time crystal may
be a good candidate for the pseudogap state in superconducting cuprates.

\begin{acknowledgments}
I thank S.I. Mukhin, B.Z. Spivak, P.A. Volkov, G.E. Volovik, and P.B.
Wiegmann for useful discussions. Financial support of Deutsche
Forschungsgemeinschaft (Project~FE~11/10-1) and of the Ministry of Science
and Higher Education of the Russian Federation in the framework of Increase
Competitiveness Program of NUST \textquotedblleft MISiS\textquotedblright
(Nr.~K2-2017-085") is greatly appreciated.
\end{acknowledgments}

\begin{widetext}

\section{Supplemental Material to `Thermodynamic quantum time crystals' by
Konstantin B. Efetov.}

\subsection{Free energy.}

\subsubsection{General scheme of the calculations.}

The free energy can be calculated minimizing the free energy functional $%
\mathcal{F}\left[ b,b_{1}\right] $, Eqs. (\ref{k10}, \ref{k6}), with respect
to $b\left( \tau \right) $ and $b_{1}\left( \tau \right) $, which leads to
Eqs. (\ref{k3}, \ref{k3a}), and calculating $\mathcal{F}\left[ b,b_{1}\right]
$ at the minimum. Apparently, exact solutions of Eqs. (\ref{k3}, \ref{k3a})
can be found only numerically, which is beyond the scope of the present
publication. As concerns the analytical study, we proceed by introducing
eigenfunctions $\Psi _{s\mathbf{p}}\left( \tau \right) $, their conjugates $%
\bar{\Psi}_{s\mathbf{p}}\left( \tau \right) ,$ and eigenenergies $\epsilon
_{s\mathbf{p}},$ satisfying equations
\begin{eqnarray}
\left( h\left( \tau ,\mathbf{p}\right) -i\Sigma _{1}b_{1}\left( \tau \right)
\right) \Psi _{s\mathbf{p}}\left( \tau \right)  &=&\epsilon _{s\mathbf{p}%
}\Psi _{s\mathbf{p}}\left( \tau \right) ,  \label{k32} \\
\bar{\Psi}_{s\mathbf{p}}\left( \tau \right) \left( \bar{h}\left( \tau ,%
\mathbf{p}\right) -i\Sigma _{1}b_{1}\left( \tau \right) \right)  &=&\epsilon
_{s\mathbf{p}}\bar{\Psi}_{s\mathbf{p}}\left( \tau \right) ,  \notag
\end{eqnarray}%
and antiperiodicity conditions
\begin{equation}
\Psi _{s\mathbf{p}}\left( \tau +1/T\right) =-\Psi _{s\mathbf{p}}\left( \tau
\right) ,\quad \bar{\Psi}_{s\mathbf{p}}\left( \tau +1/T\right) =-\bar{\Psi}%
_{s\mathbf{p}}\left( \tau \right) .  \label{k34}
\end{equation}%
Operator $h\left( \tau ,\mathbf{p}\right) $ has already been introduced in
Eq. (\ref{k6}), while its conjugate $\bar{h}\left( \tau ,\mathbf{p}\right) $
acting on the left equals
\begin{equation}
\bar{h}\left( \tau ,\mathbf{p}\right) =\left( -\overleftarrow{\partial }%
_{\tau }+\varepsilon ^{+}\left( \mathbf{p}\right) -\varepsilon ^{-}\left(
\mathbf{p}\right) \Sigma _{2}-b\left( \tau \right) \Sigma _{3}\right) .
\label{k33}
\end{equation}%
One can introduce a scalar product $\left( ,\right) $ and build an
orthonormal set of the eigenfunctions
\begin{equation}
\left( \bar{\Psi}_{s\mathbf{p}},\Psi _{s^{\prime }\mathbf{p}}\right) \equiv
T\int_{0}^{1/T}\bar{\Psi}_{s\mathbf{p}}\left( \tau \right) \Psi _{s^{\prime }%
\mathbf{p}}\left( \tau \right) d\tau =\delta _{ss^{\prime }}.  \label{k37}
\end{equation}%
Then, one can write the `electronic'\ part $\mathcal{F}_{\mathrm{el}}$
(first term in the integrand in Eq. (\ref{k10})) in the form
\begin{equation}
\frac{\mathcal{F}_{\mathrm{el}}\left[ b\left( \tau \right) ,b_{1}\left( \tau
\right) \right] }{VT}=-2\sum_{s}\int \left[ \int_{0}^{1/T}\ln \frac{\epsilon
_{s\mathbf{p}}}{T}d\tau \right] \frac{d\mathbf{p}}{\left( 2\pi \right) ^{2}}.
\label{k38}
\end{equation}

One should keep in mind that the eigenenergies $\epsilon _{s\mathbf{p}}$ are
functionals of the functions $b\left( \tau \right) $ and $b_{1}\left( \tau
\right) .$ The fact that the functional $\mathcal{F}_{\mathrm{el}}\left[
b\left( \tau \right) ,b_{1}\left( \tau \right) \right] $ can be expressed in
terms of only the eigenvalues simplifies calculations. We cannot find $%
\epsilon _{s\mathbf{p}}$ and $\Psi _{s\mathbf{p}}\left( \tau \right) $
exactly for arbitrary $b_{1}\left( \tau \right) $ and use a perturbation
theory for the eigenvalues $\epsilon _{s\mathbf{p}}$. In the zero
approximation we put $b_{1}\left( \tau \right) =0$ and find the minimum of
the functional
\begin{equation}
\mathcal{F}\left[ b\left( \tau \right) ,0\right] =\mathcal{F}_{\mathrm{el}}%
\left[ b\left( \tau \right) ,0\right] +V\int_{0}^{1/T}\frac{b^{2}\left( \tau
\right) }{2U_{0}}d\tau ,  \label{k39}
\end{equation}%
which leads to equation (\ref{k10b}). The latter can be written in the form%
\begin{equation}
b\left( \tau \right) =-U_{0}\sum_{s}\frac{\bar{\Psi}_{s\mathbf{p}}^{\left(
0\right) }\left( \tau \right) \Sigma _{3}\Psi _{s\mathbf{p}}^{\left(
0\right) }\left( \tau \right) }{\epsilon _{s\mathbf{p}}^{\left( 0\right) }},
\label{k40}
\end{equation}%
with $\Psi _{s\mathbf{p}}^{\left( 0\right) }\left( \tau \right) $ and $%
\epsilon _{s\mathbf{p}}^{\left( 0\right) }$ equal to the eigenfunctions $%
\Psi _{s\mathbf{p}}\left( \tau \right) $ and eigenvalues $\epsilon _{s%
\mathbf{p}}$ taken at $b_{1}\left( \tau \right) =0.$ Solving equations (\ref%
{k40}, \ref{k32}) is a non-trivial task even at $b_{1}\left( \tau \right)
=0. $ Nevertheless, one can find in this limit $\tau $-dependent solutions
exactly, which allows one to calculate $\mathcal{F}\left[ b\left( \tau
\right) ,0\right] $ by inserting the eigenvalues $\epsilon _{s\mathbf{p}%
}^{\left( 0\right) }$ and the solution $b\left( \tau \right) $ into Eq. (\ref%
{k38}).

As the next step, we assume non-zero $b_{1}\left( \tau \right) $, and write
\begin{equation}
b\left( \tau \right) =b_{0}\left( \tau \right) +\delta b\left( \tau \right) ,
\label{k41}
\end{equation}%
where $b_{0}\left( \tau \right) $ is the solution of Eqs. (\ref{k40}, \ref%
{k32}), and expand $\mathcal{F}_{\mathrm{el}}\left[ b\left( \tau \right)
,b_{1}\left( \tau \right) \right] ,$ Eq. (\ref{k38}), in $b_{1}\left( \tau
\right) $ and $\delta b\left( \tau \right) $ up to the second order in these
variables. This allows us to obtain the interaction between the fields $%
b\left( \tau \right) $ and $b_{1}\left( \tau \right) ,$ Eq. (\ref{k11}), and
take into account a screening of this interaction. The calculation of the
free energy functional $\mathcal{F}_{\mathrm{el}}\left[ b\left( \tau \right)
,b_{1}\left( \tau \right) \right] $ is done by substituting
\begin{equation}
\epsilon _{s\mathbf{p}}=\epsilon _{s\mathbf{p}}^{\left( 0\right) }+\epsilon
_{s\mathbf{p}}^{\left( 1\right) }+\epsilon _{s\mathbf{p}}^{\left( 2\right) },
\label{k42}
\end{equation}%
into Eq. (\ref{k38}) and calculating $\epsilon _{s\mathbf{p}}^{\left(
1\right) }$ and $\epsilon _{s\mathbf{p}}^{\left( 2\right) }$ with the help
of standard quantum-mechanical formulas%
\begin{equation}
\epsilon _{s\mathbf{p}}^{\left( 1\right) }=-\int_{0}^{1/T}\Pi _{ss}\left(
\tau \right) d\tau ,\quad \epsilon _{\mathbf{p}}^{\left( 2\right)
}=\sum_{s^{\prime }\neq s}\int_{0}^{1/T}\frac{\Pi _{ss^{\prime }}\left( \tau
\right) \Pi _{s^{\prime }s}\left( \tau \right) }{\epsilon _{s\mathbf{p}%
}^{\left( 0\right) }-\epsilon _{s^{\prime }\mathbf{p}}^{\left( 0\right) }}%
d\tau ,  \label{k43}
\end{equation}%
where
\begin{equation*}
\Pi _{ss^{\prime }}\left( \tau \right) =\bar{\Psi}_{s\mathbf{p}}^{\left(
0\right) }\left( \tau \right) \left( ib_{1}\left( \tau \right) \Sigma
_{1}+\delta b\left( \tau \right) \Sigma _{3}\right) \Psi _{s^{\prime }%
\mathbf{p}}^{\left( 0\right) }\left( \tau \right) .
\end{equation*}

As soon as the electronic part is calculated, one should minimize $\mathcal{F%
}\left[ b\left( \tau \right) ,b_{1}\left( \tau \right) \right] ,$ Eq. (\ref%
{k10}), with respect to $b_{1}\left( \tau \right) $ and $\delta b\left( \tau
\right) ,$ and calculate the free energy in terms of the solution $%
b_{0}\left( \tau \right) $ of Eq. (\ref{k40}).

\subsubsection{Unperturbed eigenfunctions $\Psi _{s\mathbf{p}}^{\left(
0\right) }\left( \protect\tau \right) $ and eigenenergies $\protect\epsilon %
_{s\mathbf{p}}^{\left( 0\right) }$.}

Following the proposed scheme we start our calculations by solving Eqs. (\ref%
{k32}, \ref{k40}). In order to avoid complicated mathematics one can simply
guess the solution $b_{0}\left( \tau \right) $ in the form of Eq. (\ref{k12}%
). This type of solutions has been used long ago for one dimensional models
of polymers \cite{brazovskii,machida,mertsching}, and more recently in Refs.
\cite{mukhin,mukhin1,mukhin2} for searching imaginary-time-dependent
solutions for order parameter.

Function $b_{0}\left( \tau \right) $, Eq. (\ref{k12}), satisfies the
following equation
\begin{equation}
\dot{b}_{0}^{2}\left( \tau \right) =b_{0}^{4}\left( \tau \right) -\gamma
^{2}\left( 1+k^{2}\right) b_{0}^{2}\left( \tau \right) +\gamma ^{4}k^{2}.
\label{k46}
\end{equation}%
At $b_{1}\left( \tau \right) =0$ we have instead of Eq. (\ref{k32}) somewhat
simpler equations
\begin{eqnarray}
&&\left( \partial _{\tau }+\varepsilon ^{+}\left( \mathbf{p}\right)
-\varepsilon ^{-}\left( \mathbf{p}\right) \Sigma _{2}-b_{0}\left( \tau
\right) \Sigma _{3}\right) \Psi _{s\mathbf{p}}^{\left( 0\right) }\left( \tau
\right) =\epsilon _{s\mathbf{p}}^{\left( 0\right) }\Psi _{s\mathbf{p}%
}^{\left( 0\right) }\left( \tau \right) ,  \notag \\
&&\bar{\Psi}_{s\mathbf{p}}^{\left( 0\right) }\left( \tau \right) \left( -%
\overleftarrow{\partial }_{\tau }+\varepsilon ^{+}\left( \mathbf{p}\right)
-\varepsilon ^{-}\left( \mathbf{p}\right) \Sigma _{2}-b_{0}\left( \tau
\right) \Sigma _{3}\right) =\epsilon _{s\mathbf{p}}^{\left( 0\right) }\bar{%
\Psi}_{s\mathbf{p}}^{\left( 0\right) }\left( \tau \right) .  \label{k47}
\end{eqnarray}%
Solutions $\Psi _{mn\mathbf{p}}^{\left( 0\right) }\left( \tau \right) $, $%
\bar{\Psi}_{mn\mathbf{p}}^{\left( 0\right) }\left( \tau \right) $ of Eqs. (%
\ref{k47}) and the eigenvalues $\epsilon _{s\mathbf{p}}^{\left( 0\right) }$
have been found exactly and further used for calculation of $\epsilon _{s%
\mathbf{p}}$ with the help of Eqs. (\ref{k43}) in Ref. \cite{efetovPRB}.

\subsubsection{Final formulas for the free energy.}

Using the eigenfunctions $\Psi _{s\mathbf{p}}^{\left( 0\right) }\left( \tau
\right) $ and eigenenergies $\epsilon _{\mathbf{p}}^{\left( 0\right) }$
obtained in Ref. \cite{efetovPRB} one can reduce Eqs. (\ref{k10b}, \ref{k40}%
) to a simpler form
\begin{eqnarray}
&&1=\frac{U_{0}}{2}\int \Big[\tanh \frac{\kappa _{\mathbf{p}}+\varepsilon _{%
\mathbf{p}}^{+}/T}{2}+\tanh \frac{\kappa _{\mathbf{p}}-\varepsilon _{\mathbf{%
p}}^{+}/T}{2}\Big]  \label{k62} \\
&&\times \frac{\left\vert \varepsilon ^{-}\left( \mathbf{p}\right)
\right\vert }{\sqrt{\left( \left( \varepsilon ^{-}\left( \mathbf{p}\right)
\right) ^{2}+\gamma ^{2}\frac{\left( 1-k\right) ^{2}}{4}\right) \left(
\left( \varepsilon ^{-}\left( \mathbf{p}\right) \right) ^{2}+\gamma ^{2}%
\frac{\left( 1+k\right) ^{2}}{4}\right) }}\frac{d\mathbf{p}}{\left( 2\pi
\right) ^{2}},  \notag
\end{eqnarray}%
which means that the function $b_{0}\left( \tau \right) ,$ Eq. (\ref{k12}),
is an exact solution of Eqs. (\ref{k10b}, \ref{k40}). In fact, there can be
many solutions of Eq. (\ref{k62}) because it contains two unknown parameters
$\gamma $ and $k.$ Using the periodicity, Eq. (\ref{k12a}), one can
determine the modulus $k$ for a given integer $m$. We concentrate here on
the limit of low temperatures $T.$ In the limit $T\rightarrow 0$ and $%
k\rightarrow 1,$ Eq. (\ref{k62}) simplifies to Eq. (\ref{k21}).

The energy $F_{\mathrm{inst}}$ entering Eq. (\ref{k14}) with the subsequent
equation has been calculated using Eqs. (\ref{k38}, \ref{k10}). In the zero
approximation in $b_{1}\left( \tau \right) $ it takes the form
\begin{equation}
\frac{F_{\mathrm{inst}}}{V}=-2\sum_{s}\ln \frac{\epsilon _{s\mathbf{p}%
}^{\left( 0\right) }}{T}\frac{d\mathbf{p}}{\left( 2\pi \right) ^{2}}+\frac{T%
}{\tilde{U}_{\mathrm{0}}}\int_{0}^{1/T}b_{0}^{2}\left( \tau \right) d\tau .
\label{k63}
\end{equation}%
Then, one should sum over all $s$ numerating eigenstates of Eq. (\ref{k47})
in the first term and integrate over $\tau $ in the second one. The energy $%
F_{\mathrm{II}}$ is obtained from Eq. (\ref{k13a}) integrating over $\tau $.
Although both the contributions can be calculated exactly at arbitrary $T$
and $k,$ only simplified formulas obtained in the limit of small $T$ and $%
k\rightarrow 1$ are displayed here in two equations following Eq. (\ref{k14}%
).

The free energy functional $\Delta \mathcal{F}\left[ b_{1},\delta b\right] $
containing both linear and quadratic terms in $b_{1}\left( \tau \right) $
and $\delta b\left( \tau \right) $ can be calculated using Eqs. (\ref{k10}, %
\ref{k38}, \ref{k42}). It can be represented in the form
\begin{equation}
\Delta \mathcal{F}\left[ b_{1},\delta b\right] =F_{\mathrm{inst}}+\mathcal{F}%
_{\mathrm{int}}\left[ b_{1}\right] +\mathcal{F}_{\mathrm{2}}\left[
b_{1},\delta b\right]  \label{k67b}
\end{equation}%
with $F_{\mathrm{inst}}$, Eq. (\ref{k63}), the linear term $\mathcal{F}_{%
\mathrm{int}}\left[ b_{1}\right] $, Eq. (\ref{k11}), and a quadratic form $%
\mathcal{F}_{\mathrm{2}}$ of $b_{1}\left( \tau \right) $ and $\delta b\left(
\tau \right) $ that can be reduced to the form,
\begin{eqnarray}
\frac{\mathcal{F}_{\mathrm{2}}\left[ b_{1},\delta b\right] }{VT}
&=&\int_{0}^{1/T}\Big[\left( \mathcal{A}\left( \tau \right) -\frac{\mathcal{C%
}^{2}}{4\mathcal{B}}\dot{b}_{0}^{2}\left( \tau \right) \right)
b_{1}^{2}\left( \tau \right) +\mathcal{B}b_{0}^{2}\left( \tau \right) \left(
\delta b\left( \tau \right) -\frac{\mathcal{C}\dot{b}_{0}\left( \tau \right)
}{2\mathcal{B}b_{0}\left( \tau \right) }b_{1}\left( \tau \right) \right) ^{2}%
\Big]d\tau ,  \notag \\
&&  \label{k67a}
\end{eqnarray}%
where
\begin{equation}
\mathcal{A}\left( \tau \right) =\mathcal{A}_{0}+\frac{\mathcal{A}_{1}}{4}%
\dot{b}_{0}^{2}\left( \tau \right) ,  \label{k71}
\end{equation}%
and the constants $\mathcal{A}_{0}$, $\mathcal{A}_{1}$, $\mathcal{B}$ and $%
\mathcal{C}$ equal
\begin{equation}
\mathcal{A}_{0}\mathcal{=}\left( 1+\frac{U_{0}}{\tilde{U}_{0}}\right) \int
\frac{1}{\sqrt{\left( \varepsilon ^{-}\left( \mathbf{p}\right) \right)
^{2}+\gamma ^{2}}}\frac{d\mathbf{p}}{\left( 2\pi \right) ^{2}},\quad
\mathcal{B=}\frac{1}{\left( \left( \varepsilon ^{-}\left( \mathbf{p}\right)
\right) ^{2}+\gamma ^{2}\right) ^{3/2}}\frac{d\mathbf{p}}{\left( 2\pi
\right) ^{2}},  \label{k69}
\end{equation}%
\begin{equation}
\mathcal{A}_{1}=\int \frac{1}{\left( \left( \varepsilon ^{-}\left( \mathbf{p}%
\right) \right) ^{2}+\gamma ^{2}\right) ^{3/2}\left( \left( \varepsilon
^{-}\left( \mathbf{p}\right) \right) ^{2}+\gamma ^{2}\frac{\left( 1-k\right)
^{2}}{4}\right) }\frac{d\mathbf{p}}{\left( 2\pi \right) ^{2}},  \label{k70}
\end{equation}%
\begin{equation}
\mathcal{C=}\int \frac{\varepsilon ^{-}\left( \mathbf{p}\right) }{\left(
\left( \varepsilon ^{-}\left( \mathbf{p}\right) \right) ^{2}+\gamma
^{2}\right) ^{3/2}\left( \left( \varepsilon ^{-}\left( \mathbf{p}\right)
\right) ^{2}+\frac{\gamma ^{2}\left( 1-k\right) ^{2}}{4}\right) }\frac{d%
\mathbf{p}}{\left( 2\pi \right) ^{2}}.  \label{k72}
\end{equation}%
The minimum of $\mathcal{F}_{\mathrm{2}}\left[ b_{1},\delta b\right] $ with
respect to $\delta b\left( \tau \right) $ is achieved at%
\begin{equation}
\delta b\left( \tau \right) =\frac{\mathcal{C}}{2\mathcal{B}b_{0}\left( \tau
\right) }\dot{b}_{0}\left( \tau \right) b_{1}\left( \tau \right) .
\label{k67c}
\end{equation}%
Then, one finds the minimum value of $\Delta \mathcal{F}\left[ b_{1},\delta b%
\right] $ (\ref{k67b}) leading to the energy $\Delta F$
\begin{equation}
\frac{\Delta F}{VT}=\frac{F_{\mathrm{inst}}}{VT}-J^{2}\int_{0}^{1/T}\left[
\mathcal{A}_{0}+\frac{1}{4}\left( \mathcal{A}_{1}\mathcal{-}\frac{\mathcal{C}%
^{2}}{\mathcal{B}_{0}}\right) \dot{b}_{0}^{2}\left( \tau \right) \right]
^{-1}\dot{b}_{0}^{2}\left( \tau \right) d\tau .  \label{k68}
\end{equation}%
In Eq. (\ref{k68}) the constant $J$ entering Eq. (\ref{k11}) is given in the
limit of low temperatures by the integral
\begin{equation}
J=\frac{1}{2}\int \frac{sgn\left( \varepsilon ^{-}\left( \mathbf{p}\right)
\right) }{\sqrt{\left( \left( \varepsilon ^{-}\left( \mathbf{p}\right)
\right) ^{2}+\gamma ^{2}\frac{\left( 1-k\right) ^{2}}{4}\right) \left(
\left( \varepsilon ^{-}\left( \mathbf{p}\right) \right) ^{2}+\gamma
^{2}\right) }}\frac{d\mathbf{p}}{\left( 2\pi \right) ^{2}}.  \label{k67}
\end{equation}

The main contribution to the integral (\ref{k68}) comes from the vicinity of
zeros of $b_{0}\left( \tau \right) $. Therefore the integral as well as $F_{%
\mathrm{inst}}$ is proportional to $2m,$ and one can integrate over the half
period of the function $b_{0}\left( \tau \right) .$ The free energy $\Delta
F $ is also proportional to $2m$ and one can calculate the energy per one
instanton replacing $b_{0}\left( \tau \right) $ by $\gamma \tanh \gamma \tau
$ and integrating over $\tau $ from $-\infty $ to $\infty .$

Equation (\ref{k68}) can be further simplified introducing a new variable of
integration $v=\gamma \tanh \gamma \tau .$ In order to compute the energy $%
\Delta F$ explicitly one should choose a specific form of the electron
spectrum, and an option used here is introduced at the bottom of the p. 3 of
the main text. Fig. 1 describes results of the calculations using this
spectrum.

\subsection{Real-time correlation functions.}

Of a special interest is a two-times correlation function $N\left( t\right) $%
, Eq. (\ref{k17}) of two current operators taken at different real times $%
t_{1}$, $t_{2},$ $t=t_{1}-t_{2}$ and oscillating in space. For the
spin-fermion model with the overlapping hot spots, the Fourier transform of
this correlation function is proportional to the cross-section of inelastic
neutron scattering at the wave vector $\left( \pi ,\pi \right) $. Generally,
non-decaying $N\left( t\right) $ can be considered as the long-range order
of the time crystal.

Starting with Hamiltonian $\hat{H}$, Eq. (\ref{k1}), and assuming that $T=0$
one can write the correlation function $N\left( t\right) ,$ Eq. (\ref{k17}),
of currents at different times $t_{1},$ $t_{2}$ and oscillating in space
with the vector $\left( \pi ,\pi \right) $ in the form (actually, the
correlation function of the currents is proportional to $N\left( t\right) ,$
and one should write a proper coefficient for comparison with experiments).
In Eq. (\ref{k17}) the angular brackets $\left\langle ...\right\rangle $
stand for quantum-mechanical averaging over the ground state of the
Hamiltonian (\ref{k1}), and
\begin{equation}
c_{p}\left( t\right) =e^{i\hat{H}t}c_{p}e^{-i\hat{H}t},\quad c_{p}^{+}\left(
t\right) =e^{i\hat{H}t}c_{p}^{+}e^{-i\hat{H}t}.  \label{k81}
\end{equation}%
This is a standard definition of the current correlations in the system in
thermodynamic equilibrium. The equivalent functional integration is based on
averaging with action $\tilde{S}$ written in real time. This action $\tilde{S%
}\left[ \chi \right] $ can be written in real time $t$ in the form%
\begin{equation}
\tilde{S}_{\mathrm{0}}\left[ \chi \right] =\tilde{S}_{\mathrm{0}}\left[ \chi %
\right] +\tilde{S}_{\mathrm{int}}\left[ \chi \right] ,  \label{k91}
\end{equation}%
where
\begin{equation}
\tilde{S}_{\mathrm{0}}\left[ \chi \right] =\sum_{p}\int_{-\infty }^{\infty
}\chi _{p}^{+}\left( t\right) \left( -i\partial _{t}+\varepsilon ^{+}\left(
\mathbf{p}\right) -\varepsilon ^{-}\left( \mathbf{p}\right) \Sigma
_{2}\right) \chi _{p}\left( t\right) dt,  \label{k92}
\end{equation}%
and%
\begin{eqnarray}
&&\tilde{S}_{\mathrm{int}}\left[ \chi \right] =-\frac{1}{4V}\int_{-\infty
}^{\infty }\Big[U_{\mathrm{0}}\Big(\sum_{p}\chi _{p}^{+}\left( t\right)
\Sigma _{3}\chi _{p}\left( t\right) \Big)^{2}  \notag \\
&&-\tilde{U}_{\mathrm{0}}\Big(\sum_{p}\chi _{p}^{+}\left( t\right) \Sigma
_{1}\chi _{p}\left( t\right) \Big)^{2}\Big]dt.  \label{k93}
\end{eqnarray}%
Using Eqs. (\ref{k93}) one can write at $T=0$ the following alternative
average instead of Eq. (\ref{k17})
\begin{equation}
N\left( t\right) =\frac{U_{0}^{2}}{V^{2}}\sum_{\mathbf{p,p}^{\prime }\mathbf{%
,}\alpha ,\alpha ^{\prime }}\left\langle \left( \chi _{p}^{+}\left( t\right)
\Sigma _{3}\chi _{p}\left( t\right) \right) \left( \chi _{p^{\prime
}}^{+}\left( 0\right) \Sigma _{3}\chi _{p^{\prime }}\left( 0\right) \right)
\right\rangle _{\tilde{S}}.  \label{k82}
\end{equation}%
In Eq. (\ref{k82}) the angular brackets denote the following average%
\begin{equation}
\left\langle ...\right\rangle _{\tilde{S}}=\frac{\int \left( ...\right) e^{-i%
\tilde{S}\left[ \chi \right] }D\chi }{\int e^{-i\tilde{S}\left[ \chi \right]
}D\chi }.  \label{k94}
\end{equation}%
(A rotation in the space of the band numbers $1,2$ has been made when
passing from Eq. (\ref{k17}) to Eqs. (\ref{k93}, \ref{k82})). In order to
calculate the average in Eq. (\ref{k94}) we decouple the interaction with
the help of the Hubbard-Stratonovich transformation using auxiliary real $%
B\left( t\right) $ and $B_{1}\left( t\right) $ fields. This leads us to the
electron part of action $\mathcal{S}\left[ \chi \,\chi ^{+},B,B_{1}\left(
t\right) \right] $ containing both fermionic $\chi ,\chi ^{+}$ and bosonic $%
B\left( t\right) ,$ $B_{1}\left( t\right) $ fields%
\begin{equation}
\mathcal{S}\left[ \chi \,\chi ^{+},B,B_{1}\left( t\right) \right]
=\int_{-\infty }^{\infty }\chi _{p}^{+}\left( t\right) \mathcal{H}\left( t,%
\mathbf{p}\right) \chi _{p}\left( t\right) dt  \label{k94a}
\end{equation}%
with the operator $\mathcal{H}\left( t,\mathbf{p}\right) $ equal to
\begin{equation}
\mathcal{H}\left( t,\mathbf{p}\right) =-i\partial _{t}+\varepsilon
^{+}\left( \mathbf{p}\right) -\varepsilon ^{-}\left( \mathbf{p}\right)
\Sigma _{2}-i\left( B\left( t\right) \Sigma _{3}+B_{1}\left( t\right) \Sigma
_{1}\right) .  \label{k97}
\end{equation}%
Then, we integrate over $\chi $ and reduce the full action $S$ to the form

\begin{equation}
S=-\ln \left[ \int \exp \left[ -i\mathcal{S}\left[ B,B_{1}\right] \right]
DBDB_{1}\right] ,  \label{k95}
\end{equation}%
where the action functional $\mathcal{S}\left[ B,B_{1}\right] $ equals
\begin{eqnarray}
\mathcal{S}\left[ B,B_{1}\right]  &=&\int_{-\infty -}^{\infty }\Big[-2\sum_{%
\mathbf{p}}\mathrm{tr}\left[ \ln \left( \mathcal{H}\left( \tau ,\mathbf{p}%
\right) \right) \right] _{t,t}  \label{k96} \\
&&-V\left( \frac{B^{2}\left( t\right) }{U_{\mathrm{0}}}-\frac{%
B_{1}^{2}\left( t\right) }{\tilde{U}_{\mathrm{0}}}\right) \Big]dt,  \notag
\end{eqnarray}%
Minimizing $\mathcal{S}\left[ B,B_{1}\right] $ with respect to $B\left(
t\right) $ and $B_{1}\left( t\right) $ we come to equations
\begin{eqnarray}
B\left( t\right)  &=&iU_{\mathrm{0}}\mathrm{tr}\int \Sigma _{3}\left[
\mathcal{H}^{-1}\left( t,\mathbf{p}\right) \right] _{t,t}\frac{d\mathbf{p}}{%
\left( 2\pi \right) ^{2}},  \label{k98} \\
B_{1}\left( t\right)  &=&-i\tilde{U}_{\mathrm{0}}\mathrm{tr}\int \Sigma _{1}%
\left[ \mathcal{H}^{-1}\left( t,\mathbf{p}\right) \right] _{t,t}\frac{d%
\mathbf{p}}{\left( 2\pi \right) ^{2}}.  \label{k98a}
\end{eqnarray}%
Comparing Eqs. (\ref{k98}, \ref{k98a}) with Eqs. (\ref{k3}, \ref{k3a}) we
come to conclusion that the functions $B\left( t\right) $ and $B_{1}\left(
t\right) $ are related to $b\left( it\right) $ and $b_{1}\left( it\right) $
by Eqs. (\ref{k15}).

It is very important that if $B\left( t\right) $ and $B_{1}\left( t\right) $
are solutions of Eqs. (\ref{k98}, \ref{k98a}), then $B\left( t-t_{0}\right) $
and $B_{1}\left( t-t_{0}\right) $ are also solutions at an arbitrary $t_{0}$%
. It is also clear that there can be many solutions even at a fixed $t_{0}$.
For example, for $B_{1}\left( t\right) =0$ one comes to Eq. (\ref{k62}) for
any period of the function $B\left( t\right) $ given by Eq. (\ref{k16}). The
relations (\ref{k15}) allow one to obtain proper $B\left( t\right) $ and $%
B_{1}\left( t\right) $ as soon as $b\left( \tau \right) $ and $b_{1}\left(
\tau \right) $ are obtained from the condition for the minimum of the free
energy functional $\mathcal{F}\left[ b,b_{1}\right] $, Eq. (\ref{k10}).

Now, using Eq. (\ref{k94a}, \ref{k97}) we can integrate over the fermionic
fields $\chi ,\chi ^{+}$ in Eq. (\ref{k82}) to obtain in the limit $%
V\rightarrow \infty $
\begin{equation}
N\left( t_{1}-t_{2}\right) =-U_{\mathrm{0}}^{2}\overline{\int \mathrm{tr}%
\left[ \Sigma _{3}\mathcal{H}^{-1}\left( t_{1}-t_{0},\mathbf{p}_{1}\right) %
\right] \frac{d\mathbf{p}_{1}}{\left( 2\pi \right) ^{2}}\int \mathrm{tr}%
\left[ \Sigma _{3}\mathcal{H}^{-1}\left( t_{2}-t_{0},\mathbf{p}_{2}\right) %
\right] \frac{d\mathbf{p}_{2}}{\left( 2\pi \right) ^{2}}},  \label{k101}
\end{equation}%
where the bar stands for the averaging over the period of the structure.
Integration over $t_{0}$ is absolutely necessary because the extremum of the
action functional is degenerate with the respect to the time shifts, and one
should integrate over all the extremum states. Finally, using Eq. (\ref{k98}%
) we come to Eqs. (\ref{k18}, \ref{k16}).
\end{widetext}

\end{document}